\documentclass[useAMS,usenatbib]{mn2e}
\usepackage{graphicx}

\title[Finding benchmark brown dwarfs]
{Finding benchmark brown dwarfs to probe the sub-stellar IMF as a function of time.}

\author[Pinfield D. J., et al.]{
D. J. Pinfield$^1$\thanks{E-mail: dpi@star.herts.ac.uk}, 
H. R. A. Jones$^1$, P. W. Lucas$^1$, T. R. Kendall$^1$, S. L. Folkes$^1$, \\
\newauthor A. C. Day-Jones$^1$, R. J. Chappelle$^2$ and I. A. Steele$^3$ \\
$^1$Centre for Astrophysics Research, Science \& Technology Research Institute, 
Department of Physics Astronomy \& Mathematics, \\
University of Hertfordshire, College Lane, Hatfield, Hertfordshire, AL10 9AB, UK \\
$^2$Astronomical Institute, Academy of Sciences of the Czech Republic, 
Bocni II/1401a, 141 31  Prague, Czech Republic \\
$^3$Astrophysics Research Institute, Liverpool John Moores University,
  Twelve Quays House, Egerton Wharf, Birkenhead, CH41 1LD
}

\begin{document}

\date{Received in original form July 2005}

\pagerange{\pageref{firstpage}--\pageref{lastpage}} \pubyear{2005}

\maketitle

\label{firstpage}

\begin{abstract}
Using a simulated disk brown dwarf (BD) population, we find that new large area infrared surveys are expected 
to identify enough BDs covering wide enough mass--age ranges to potentially measure the 
present day mass function down to $\sim$0.03M$_{\odot}$, and the BD formation history out to 10 Gyr, at 
a level that will be capable of establishing if BD formation follows star formation. We suggest these 
capabilities are best realised by spectroscopic calibration of BD properties ($T_{\rm eff}$, $g$ and [M/H]) 
which, when combined with a measured luminosity and an evolutionary model can give BD mass and age 
relatively independent of BD atmosphere models. Such calibration requires an empirical understanding 
of how BD spectra are affected by variations in these properties, and thus the identification and study 
of ``benchmark BDs'' whose age and composition can be established independently.

We identify the best sources of benchmark BDs as young open cluster members, moving group members, 
and wide ($>$1000AU) BD companions to both subgiant stars and high mass white dwarfs (WDs). To accurately 
asses the likely number of wide companion BDs available we have constrained the wide L dwarf companion 
fraction using the 2MASS All Sky Survey, and find a companion fraction of 2.7$^{+0.7}_{-0.5}$\% for 
separations of $\sim$1000--5000AU. This equates to a BD companion fraction of 34$^{+9}_{-6}$\% if one 
assumes an $\alpha\sim$1 companion mass function. Using this BD companion fraction we simulate populations 
of wide BD binaries, and estimate that 80$^{+21}_{-14}$ subgiant--BD binaries, and 50$^{+13}_{-10}$ 
benchmark WD--BD binaries could be identified using current and new facilities. The WD--BD binaries should 
all be identifiable using the Large Area Survey component of the UKIRT Infrared Deep Sky Survey, combined 
with the Sloan Digital Sky Survey. Discovery of the subgiant--BD binaries will require a NIR imaging campaign 
around a large ($\sim$900) sample of Hipparcos subgiants. If identified, spectral studies of these benchmark 
brown dwarf populations could reveal the spectral sensitivities across the $T_{\rm eff}$, $g$ and [M/H] 
space probed by new surveys.
\end{abstract}


\section{Introduction}

\subsection{The IMF and formation history}

The initial mass function (IMF) and formation history are the most important observational results 
produced by the process of Galactic star formation. These two observational cornerstones provide 
a test-bed for our theoretical understanding of this process. Brown dwarfs (BDs; mass$<$0.075M$_{\odot}$) 
populate the lowest mass extreme of the IMF, and current theory suggests that the form of the IMF 
in this mass range could be particularly sensitive to the initial conditions prevalent for low-mass 
objects. For instance, \citet{delgado04} predicts a higher fraction of BDs for a shallower initial 
slope of the turbulent velocity spectrum in the formation cloud. \citet{bate05} suggests that in denser 
clouds, with lower mean thermal Jeans mass, more BDs will form relative to stars. \citet{chabrier03} 
presents observational evidence suggesting a possible increase in the characteristic mass of star 
formation that decreases over time, between conditions characteristic of the spheroid (or thick disk) 
to present-day conditions. Alternatively, \citet{ashman90} describes how significant populations of 
halo (or thick disk) BDs may have formed in cooling flows, and discusses the possibility of a pressure 
dependent IMF.

One can measure the BD IMF in open clusters as well as younger pre-main sequence clusters emerging 
from their nascent clouds. However, \citet{lada03} found that the birth rate of embedded clusters 
exceeds that of visible open clusters by an order of magnitude or more, and that less than 4--7\% 
of embedded clusters survive emergence from molecular clouds to become bound clusters. The most 
complete way, therefore, of measuring the BD IMF is from the local disk population itself. Furthermore, 
if BD formation rates are significantly affected by average gas density, pressure and metallicity ([M/H]) 
in a different way to star formation rates, then we would expect the BD formation history and [M/H] 
distribution to be sensitive to these factors, and differ in a characteristic way to the stellar 
distributions.

Measuring these distributions is a major challenge. The nature of BD evolution (cooling and 
fading with time) means that the mass-luminosity relation depends strongly on age, and one cannot 
determine either mass or age from a BD's luminosity alone. The usual approach to this problem is to 
fit synthesised $T_{\rm eff}$ and luminosity functions (constructed with different IMFs and formation 
histories) to observed BD populations \citep{allen05,deacon05}. However, this method is sensitive only 
to quite drastically different formation histories (eg. one can discriminate between a single halo burst 
of BD formation 9-10 Gyr ago and a uniform BD birth rate), and then only if one assumes a non evolving 
IMF \citep{burgasser04b}. This method also has little chance of constraining the BD [M/H] distribution.

It is a particularly important time for this field, because large scale near infrared (NIR) surveys 
(currently championed by the 2-Micron All Sky Survey -- 2MASS; \citet{skrutskie97}, and the Denis 
survey; \citet{epchtein97}), are being taken to larger telescopes, with the start of the ``UKIRT 
Infrared Deep Sky Survey'' (UKIDSS) on the UK Infrared Telescope, the capabilities of the WIRCam 
instrument on the Canada France Hawaii Telescope, and the approach of the ``Visible and Infrared
Survey Telescope for Astronomy'' (VISTA). This paper will, in part focus on UKIDSS, as this 
large scale 7000 square degree survey \citep{hambly03} has well defined parameters and sensitivities. 
It consists of several sub-surveys (varying in sky coverage and depth), the largest of which is the 
``Large Area Survey'' (LAS) covering 4000 sq degs in four NIR bands, with photometric limits $\sim$4 
magnitudes fainter than 2MASS.

\subsection{Aims and paper structure}
In light of the significant advances in survey facilities, we would like to more ambitiously 
assess the issue of how best one could measure the disk BD IMF, formation history and [M/H] 
distributions. Ideally, one needs a method to directly measure the mass, age, and composition 
of the individual BDs making up the local disk population. Disk BDs are generally very cool 
objects with dusty upper atmospheres, comprising very young M dwarfs, a subset of L dwarfs 
(2300--1300K; \citet{kirkpatrick99}) and all T dwarfs (1300--700K; \citet{burgasser99}). In 
general, to determine a BD's mass, age and composition, one must measure $T_{\rm eff}$, $g$ 
and [M/H] from spectral properties. However, this requires detailed knowledge of the spectral 
dependencies on these properties, which must come from either theoretical models or the 
spectroscopic study of a varied population of BDs with well constrained physical properties. 
Such BDs could be used as fiducial calibrators in $T_{\rm eff}$/$g$/[M/H] space, and we thus 
refer to them as {\it benchmark brown dwarfs}.

In this paper we consider the best types of benchmark BD, and estimate the number of benchmarks 
that could be found in near future surveys, as well as the range of physical properties that 
these benchmarks could have. We also compare the expected benchmark properties to those likely 
amongst the BD populations from the new surveys, and discuss the implications for calibrating 
the IMF and formation history. In Section 2 we use a simulation of the local disk low-mass 
population to define the BD mass--age range over which current and future surveys are able to 
detect enough BDs to constrain the IMF and formation history. In Section 3 we discuss 
previous work on BDs with constrained properties, the search for sensitive spectral features, 
and how one could build on this by identifying new types of benchmark BDs with better constrained 
properties. Sections 4, 5, 6 and 7 discuss these benchmark populations in more detail, and combine 
new observational results with simulations to estimate the numbers of benchmark BDs that could 
be discovered in the near future. Section 8 discusses the distribution of the predicted benchmark 
BDs in mass/age/[M/H] space, comparing these properties with the survey parameter space defined 
in Section 2. Section 9 contains our conclusions.
\section{NIR survey sensitivities to the IMF and formation history}
\subsection{A simulation of disk brown dwarfs}
In order to determine the region of the sub-stellar mass--age plane that we expect large scale 
surveys to probe, we have simulated a local disk population of BDs. We do not intend that this 
should be a particularly detailed assessment, which would be a sensitive function of the BD 
population itself (e.g. dependence on structure in the luminosity and $T_{\rm eff}$ functions; 
\citet{allen05}). Instead we consider a particular type of BD population that might be considered 
``typical'' based on current constraints (e.g. \citet{chabrier03}). We thus assumed a uniform 
spatial distribution within the plane itself (ie. uniform surface density), onto which we imposed 
an $\alpha$=1 IMF (where $\xi(m)\propto m^{-\alpha}$) and a formation history identical to the 
local stellar population \citep{rocha-pinto00}. Absolute J-band magnitudes were derived from mass 
and age, using theoretical models (currently available for solar metallicity; \citet{baraffe98,
chabrier00a,baraffe03}). We then determine the vertical height above the plane ($z$) by imposing 
an exponential $z$ distribution (extending both positive and negative) on the population, which 
we normalised (for each BD) using the relation between scale-height (H) and age from \citet{just03}. 
We thus account for the fact that disk heating causes older populations to become more vertically 
dispersed in the disk, with resulting lower number densities in the plane itself (scaling with 
1/H(age)). Distances and Galactic latitudes ($b$) were then derived trigonometrically, and apparent 
magnitudes determined accordingly. At this point we counted the number of simulated stars out to 
distances of 20 parsecs with masses from 0.09--0.1M$_{\odot}$, and normalised the number of objects 
in our simulation to produce a total of 184 in this mass and distance range, consistent with the 
observed low-mass stellar mass function \citep{reid99,chabrier01}.

\subsection{Brown dwarfs in large scale NIR surveys}

To select ``survey populations'' of BDs, we use 
the sensitivities of 2MASS and the UKIDSS LAS . We selected simulated sub-samples with J$\le$16 
and $|b|>15$ to represent 2MASS BDs, and with J$\le$19.5 and $|b|>57$ (approximating 4000 square degrees 
in the Galactic cap) to represent UKIDSS LAS BDs. We also imposed distance limits on our sub-samples. 
2MASS allows one to probe for late L dwarfs (M$_J\sim$14.5) out to 20pc, and the UKIDSS LAS will be 
sensitive to late T dwarfs (M$_J\sim$15.5) out to 50pc. Figure 1 shows the mass--age distribution of 
our simulated 2MASS ($<$20pc) and UKIDSS LAS ($<$50pc) BD samples. Approximate M, L and T 
spectral type divisions are shown as dotted lines (where class transition $T_{\rm eff}$s were estimated 
from \citet{knapp04}). In the 2MASS figure the 20pc distance limit is essentially the L/T transition. 
For the UKIDSS LAS, the 50pc distance limit corresponds to a spectral type of $\sim$T7, and is indicated 
with a dashed line. Note that for the youngest BDs ($<$ a few hundred Myr) the assumption of a uniform 
spatial distribution where scale height changes cleanly with age will not be ideal, since many very 
young objects are likely to be in super cluster structures \citep{montes01}. The choice of our particular 
theoretical models is an additional source of uncertainty.
\begin{figure}
\includegraphics[width=84mm]{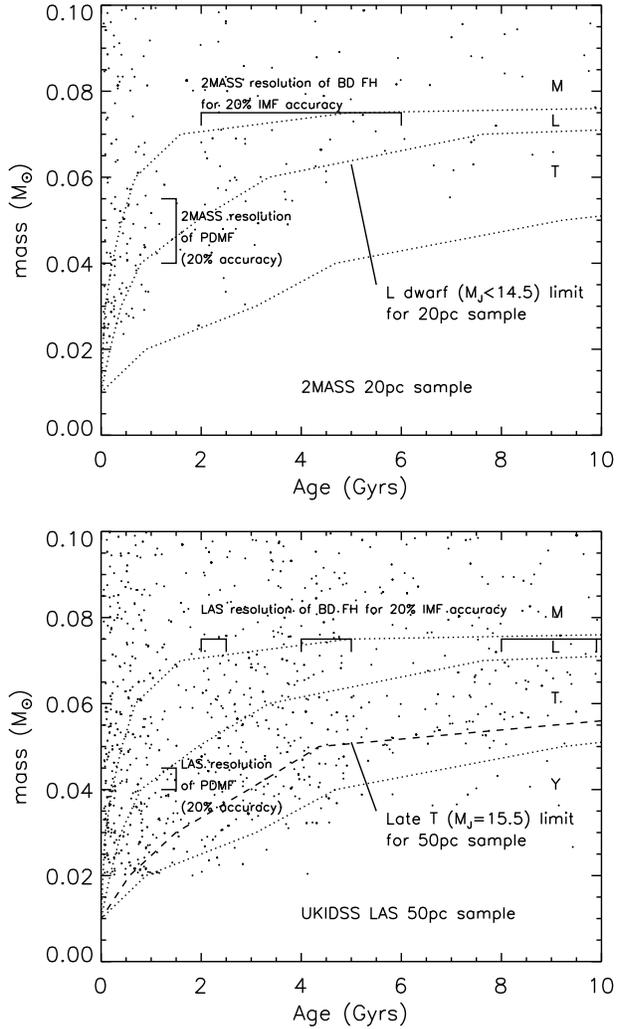}
\caption{The mass--age plane for low-mass objects in simulated 2MASS and UKIDSS 
LAS populations. The 2MASS population is an all sky 20pc sample. The UKIDSS LAS 
population is for 4000 square degrees in the Galactic cap out to 50pc. Estimated 
spectral class divisions are shown with dotted lines. The 20 and 50pc limits are 
also shown, corresponding to the L/T transition for 2MASS, and $\sim$T7 for UKIDSS 
LAS. The resolution of the BD formation history and present day mass function 
(PDMF) represent age and mass ranges in which the surveys should provide 
$\sim\pm$20\% accuracy (see text).}
\end{figure}

\subsection{The mass-age region probed by surveys}

We wish to consider the mass--age region in which we can detect enough BDs to say something specific about 
their IMF and formation history. As a particularly useful goal, we chose to address the issue of whether 
one will be able to make a useful comparison between BD formation history and star formation history 
(which we define using \citet{rocha-pinto00}). This star formation history is characterised by a series 
of formation bursts (see \citet{burgasser04b}), lasting 1--3 Gyr, and during which formation rates increase 
by $\sim$40--50\%. To measure such increases we would need an accuracy of $\pm$20\% or better in the 
number of BDs in a particular burst, thus requiring $\sim$25 (assuming Poisson statistics). We indicate 
in Figure 1 age ranges containing the required number of BDs lying above the completeness limit. We 
label these age ranges as the resolution of the BD formation history. In a similar, but somewhat more 
arbitrary way, we indicate a mass range containing $\sim$25 BDs with ages from 0.5--1.5 Gyr (where we exclude 
the very young BDs for reasons mentioned in Section 2.2) to represent the smallest mass range in which we 
might constrain the present day mass function (PDMF) with $\sim$20\% accuracy. We label this mass range 
as the resolution of the PDMF.

It can be seen that the 2MASS resolution of the formation history is rather low, and will not be able 
to resolve star-formation-like bursts of 1-3 Gyr duration. The UKIDSS LAS resolution of the formation 
history is significantly better, and should be able to constrain bursts of this type even out to the 
full disk age of $\sim$10 Gyr. The 2MASS resolution of the PDMF is sufficient to constrain its shape 
at the $\Delta$M$\sim$0.015M$_{\odot}$ level, down to $\sim$0.04M$_{\odot}$, which could be extended 
with a closer sample of T dwarfs (eg. \citet{burgasser04b}). The UKIDSS LAS has a mass resolution 3 times 
better than this, and should provide a significantly more detailed IMF down to $\sim$0.03M$_{\odot}$. 
The mass--age region of interest is thus defined by the UKIDSS LAS 50pc population, lying below the 
BD limit (0.075M$_{\odot}$) and above the dashed line T7 cut-off in Figure 1b.

\section{The use of benchmark brown dwarfs}

Previously, \citet{mcgovern04} made a spectroscopic study of a selection of youthful late M/early 
L dwarfs with ages constrained through membership of young associations (1--300Myrs). They compared 
these spectra to that of an M9.5III+ Mira variable ($\log{g}\sim$0) as well as to older disk dwarfs, 
and found a variety of surface gravity sensitive features, including VO bands (more rounded at lower 
$g$), the alkali metal lines KI, RbI, NaI and CsI (weaker at low $g$), and lithium absorption (stronger 
for low $g$) in the optical. They also discovered that in the J-band, very low $g$ causes the appearance 
of the $\phi$($\Delta\nu$=-1) band heads of TiO, the A-X ($\Delta\nu$=-1) band of VO, TiO and VO at 
1.14-1.20 microns, and variations in the strength of FeH, KI and H$_2$O features. \citet{mohanty04} has 
used some of these spectral features (TiO, NaI and KI) to fit $T_{\rm eff}$ and $g$ for 13 very young 
M5--7.5 dwarfs from the Upper Sco and Taurus star-forming regions. The $g$ values they obtain are 
consistent with isochrone predictions for the cluster members except for the two coolest objects, 
and it is clearly that our understanding of how low $g$ can affects the spectra 
of young late M BDs is improving.

However, only a small fraction of BDs have late M spectral type, and as one moves to cooler $T_{\rm eff}$s, 
BD atmospheres become more dusty. The physics of dust formation in such objects represents a large 
uncertainty in the atmospheric models of L and T dwarfs, as well as having a profound effect of the 
emergent spectra. Observationally, there is a significant spread in L and T broad band colours 
\citep{golimowski04,leggett02,cruz03} that does not correlate with changes in spectral type. \citet{knapp04} 
noted that the scatter seen in the H-K colour of a sample of T dwarfs showed a correlation with the 
strength of the KI doublet -- a feature expected to be $g$ sensitive. Some L dwarfs also stand out by 
there surprisingly blue colours, and show enhanced FeH, KI and H$_2$O. This has been interpreted as 
resulting from a high $g$ atmosphere depleted of dust due to a higher rate of ``rain-out''. In this 
interpretation the amount of dust induced reddening that an L dwarf spectrum undergoes would be highly 
sensitive to $g$. Two L dwarfs with halo kinematics also show particularly unusually blue near infrared 
colours \citep{burgasser03a,burgasser04a} . These objects also show strong FeH features, and are thought 
to be extremely metal poor halo subdwarfs, where H$_2$ absorption is depressing the H- and K-band fluxes. 
Thus, the indications are that the spectral properties of L and T dwarfs are highly sensitive to $g$ 
and [M/H], as well as to $T_{\rm eff}$.

Despite this potentially promising situation, it is not currently possible to fit L and T dwarf 
$T_{\rm eff}$, $g$ and [M/H] from spectral synthesis. Atmospheric models that assume a photospheric 
dust distribution in equilibrium with the gas phase show significant discrepancies for $T_{\rm eff}<$1800K 
($\ge$L3; e.g. \citet{leggett01}). More sophisticated treatments of dust have since provided broader 
agreement between theory and the bulk properties of L and T dwarfs, although many uncertainties remain. 
For instance, fits to theoretical model colours \citep{marley02} of the T dwarf Gl 570D \citep{knapp04} 
suggested a $T_{\rm eff}$ of $\sim$950K. However, independent constraints placed on this T dwarf from 
its companion stars indicate a $T_{\rm eff}$ of 784--824K \citep{geballe01}. Also, while the latest 
models of \citet{burrows06} show a reasonable correspondence between theoretical spectra and some 
observations, they also show a variety of discrepancies (e.g shape discrepancies in the H and K bands, 
and deviations in the Y/Z peak), and are unable to fit the colours of the latest L dwarfs.

%
%

To properly understand the sensitivity of both broad band and narrower band spectral features to 
variations in $T_{\rm eff}$, $g$ and [M/H], it seems clear that one needs to spectroscopically 
study larger samples of benchmark BDs with more tightly constrained properties. Although brighter 
benchmark targets would allow for higher signal-to-noise (SN), higher resolution studies, the 
significant broad band variations in L and T dwarf colour and band strengths can easily be 
measured at low resolution, and signal-to-noise of $\sim$20--30. Benchmark BDs could thus be 
as faint as J$\sim$20, and still be studied spectroscopically. J-band spectra (R$\sim$500, 
SNR$\sim$20) could constrain spectral features at the level of atomic lines and can be measured 
in $\sim$2hr on an 8m telescope, and even for the blue T dwarfs one should be able to measure 
K-band spectra (R$\sim$100, SNR$\sim$20) sufficient to accurately constrain the continuum shape 
in a reasonable time. It is, however, important that the benchmark objects have well known 
distances, so that their luminosities can be determined. Their age must also be well constrained 
so that their radii and in particular their mass can be estimated from evolutionary models 
(although note that L and T dwarf radii are largely, but not totally, independent of mass and 
age, within a range of $\sim$30\%). The $T_{\rm eff}$ and $g$ can thus be determined (from 
luminosity and radius, and mass and radius respectively), and if [M/H] is also known, then the 
benchmark BDs will have the full complement of atmospheric properties. Evolutionary models are 
broadly consistent with each other and observations, and it is thus highly desirable that the 
atmospheric properties derived rely on evolutionary models as opposed to atmosphere models.

The best places to find such populations of benchmarks are;

\begin{itemize}
  \item{In open clusters, whose age, composition and distance may be well known.}
  \item{As nearby members of kinematic moving groups which have known age and composition.}
  \item{In well separated multiple or binary systems, where the age, distance, and 
possibly the metallicity can be inferred through association with the companion stars.}
\end{itemize}

\section{Benchmark brown dwarfs in open clusters and moving groups}

Young benchmark BDs may be found in open clusters, where the stellar membership is well studied 
and composition and age are well constrained by e.g. measuring the upper main sequence turnoff 
\citep{sarajedini99} or the magnitude of the lithium depletion edge (\citet{stauffer98}; 
\citet{barrado04}). Provided the cluster is not too young (e.g. $<$50 Myrs), then the age spread 
(a few Myrs; e.g. \citet{belikov00,park02}) should not cause significant uncertainty in the age 
of the members. One simply has to confirm cluster membership of a BD using photometry, proper motion 
and radial velocity. The upper age limit for which cluster BD populations may no longer be easily 
probed is set by dynamical evaporation of BDs from these environments. By $\sim$600--800Myr the 
identification of cluster BDs becomes difficult. This is the age of the Hyades, in which 
\citet{dobbie02b} did not find any BDs in a 10.5\,deg$^2$ survey sensitive to $I$\,$\sim$\,20. 
Even with a substantially larger (17.4\,deg$^2$) deeper ($I$\,$\sim$\,$z$\,$\sim$24) CFHT12K survey, 
\citet{bouvier05} have identified only $\sim$2 Hyades BDs (confirmed astrometrically), and a similar 
situation is seen in Praesepe \citep{chappelle05}, a cluster of approximately the same age.

At present, the known BD membership in such regions is an excellent source of benchmark objects. Many of 
the BDs identified in open clusters (e.g. Pleiades, Praesepe, $\alpha$\,Per) have been found by deep, 
wide-field imaging \citep{bouvier98,moraux01,dobbie02a,chappelle05,barrado02}. The number of cluster BD 
benchmarks will increase significantly as the UKIDSS Galactic Cluster Survey (GCS) obtains deep imaging 
of six young cluster populations. However, the spectral sensitivity of young BDs to large metallicity 
changes is not addressed by these populations, since they all have broadly solar [M/H] (-0.1--+0.2\,dex).

\begin{figure}
\includegraphics[width=84mm]{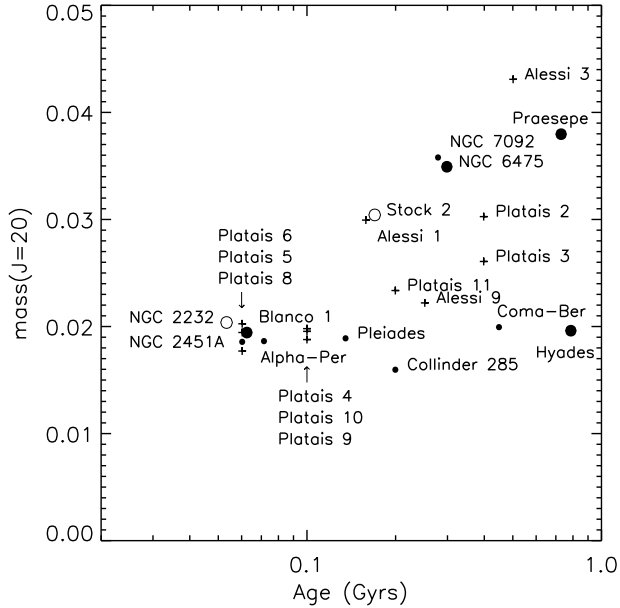}
\caption{The J=20 mass limit for open clusters from \citet{dias02}. Small filled circles 
have [M/H]=-0.1--0.1, and larger circles have [M/H]=$\pm$0.1--0.2 (filled circles are metal 
rich and open metal poor). Clusters with unknown [M/H] are shown as plus signs. Where points are 
congested, cluster names for several plus signs have been grouped together, with an arrow indicating 
the location of the group.}
\end{figure}
To address this issue further, Figure 2 shows a plot of the theoretical mass limit probed in open 
clusters to an apparent magnitude limit of J=20, verses cluster age. Clusters within 1Kpc have been 
taken from \citet{dias02}. We used the Lyon group models to calculate the mass limits, as well as 
expected T$_{\rm eff}$s for cluster members at these limits. We only plot clusters in which one could 
hope to detect L dwarfs down to $T_{\rm eff}\sim$1800K, and limit the cluster ages to $\ge$50 Myrs. 
[M/H] values from \citet{dias02} have been supplemented (and updated in some cases) by additional values 
from the literature \citep{castellani02,strobel91,cameron85,jeffries99,piatti95,claria96}. The plotting 
symbols represent these [M/H] values, where small filled circles indicate [M/H]=-0.1--0.1, larger filled 
and open circles indicate clusters that are metal-rich and poor respectively, with [M/H]=$\pm$0.1--0.2. 
Clusters without published [M/H] values are shown as plus signs.

There are 24 clusters in which one could potentially detect benchmark L and possibly T dwarfs, including 
5 of the 6 GCS clusters (IC 4665 is just below our age limit). Six of these have near solar [M/H]=-0.1--+0.1, 
four are slightly metal rich with [M/H]=0.1--0.2, two are slightly metal poor with [M/H]=-(0.2--0.1), and 
twelve have no [M/H] determination (all twelve are recently discovered clusters from the analysis of 
Hipparcos and Tycho-2 data; \citet{platais98,alessi03}). There are clearly many clusters in which one 
might detect benchmark BDs. However, we would draw attention to the fact that there are no young clusters 
with measured [M/H]$>$0.2 or $<$-0.2 in which one could detect benchmarks. This lack of known metal rich 
young clusters is of particular importance since the local young stellar population has [M/H] as high as 
0.3 dex (see \citet{edvardsson93}). It would thus be desirable to measure [M/H] of the twelve recently 
discovered clusters, to provide new cluster targets for benchmark searches.

Moving group populations are distinguishable from the field by their astrometric properties.  
A moving group remains kinematically distinct within the general field population at ages 
$<$\,1\,Gyr, before being dispersed by disk heating mechanisms (e.g. \citet{desimone04}). They 
are thought to originate in the same environment as open clusters. As progenitor gas is cleared 
by OB star winds, and the natal cluster expands, stars with sufficiently high velocities become 
unbound and form a young, coeval moving group, possibly leaving behind a bound open cluster 
\citep{kroupa01}. Before dispersal after $\sim$1\,Gyr, moving groups therefore consist of young 
populations with characteristic space motions, and membership of such a group can be used 
to accurately constrain the age and composition of a BD \citep{ribas03,pokorny04}.

Confirmation of moving group membership requires accurate space motions ($\pm$\, few km\,s$^{-1}$), 
from proper motions and radial velocities. The measurement of the proper motions of brighter objects 
should be achievable using existing data (e.g. 2MASS coupled with SuperCOSMOS). Alternatively, near 
infrared astrometric techniques are capable of centroiding at the 2mas level \citep{smart05}, and 
could be brought to bear on intrinsically fainter (late L and T dwarf) and more distant moving 
group benchmark candidates, reaching the required level of proper motion accuracy over a fairly 
short (1--2yr) base-line. Radial velocities should be measurable using NIR echelle spectrographs 
on 8m telescopes in the near future, which should provide $\sim$km/s accuracy down to J$\sim$17 
for L and T dwarfs. In addition to kinematics, moving group members with ages $<$1\,Gyr should 
possess many of the differentiative diagnostics applicable to young cluster BDs. Chief among 
these are the presence of lithium ({\it via} the 6708\AA~ line, for ages $<$100\,Myr) as well as 
numerous gravity-sensitive near-infrared spectroscopic features \citep{mcgovern04}.

As observations confirm complete, homogeneous samples of brown dwarfs in young clusters and 
moving groups, such BDs can be immediately defined as benchmark objects in the $<$1 Gyr age range.

\section{Brown dwarf companions at wide separation}

BD benchmarks do not however, have to be associated with large populations of stars. They could also 
be members of multiple or binary systems \citep{geballe01,smith03,kirkpatrick01,wilson01}. It has been 
known for some time that the degree of multiplicity amongst very young stars is greater than that of the 
more evolved field star populations \citep{leinert93,duquennoy91}, and thus that the majority of binary 
systems form together in their nascent clouds. Binary components can therefore generally be assumed to 
share the same age and composition. The usefulness of such BD companions as benchmark objects will 
depend on the separation of the components (i.e. if the BD companion can be resolved, and a spectra 
taken), and the accuracy with which one can constrain the age of the star(s) in the system.

It is known that BD companions to F-M main sequence stars are fairly common at separations $>$1000AU 
(\citet{gizis01} estimated a companion fraction of 18$\pm$14\%), in stark contrast to the lack of BD 
companions at very close separation (the so called BD desert at $<$3AU; e.g. \citet{marcy00}), as well 
as at larger separations ($<$ a few hundred AU; \citet{mccarthy04}). However, this companion fraction 
was estimated using just 3 companion BDs discovered in a fraction of the sky and confirmed by common 
proper motion. The uncertainties associated with this value are thus very large due to small number 
statistics, and we have therefore chosen to re-asses the level of the wide BD companion fraction, 
to improve these constraints.

To do this we selected all Hipparcos stars out to 50pc with $\pi/\Delta\pi>4$ and $|b|>$30 
as target primaries, and searched around these for L dwarfs using a 2MASS Gator cone search in the 
2MASS All Sky Release. For each target primary we searched out to a separation corresponding to 5000AU 
at the distance of the star, or a maximum of 300 arcseconds (the Gator limit) for stars within 16.67pc. 
We searched for sources with J-K$\ge$1.1 and J$\le$16.1, and with either no optical counterpart, or 
R-K$\ge$5.5. Sources were then visually inspected using postage-stamp image data from 2MASS, SuperCOSMOS 
(R-band and, where available I-band) and Sloan DR4 (where available), to ensure a clean photometric 
candidate sample.

The top left plot in Figure 3 shows our candidates on an M$_J$, J-K colour-magnitude diagram (CMD), 
where in each case we have assumed the candidate to be at the distance of its associated primary star. 
To identify candidates whose colour and magnitude are inconsistent with companionship, we have defined 
a region (the dashed box) where we expect L dwarfs to lie, using the L dwarf M$_J$ range from 
\citet{knapp04} coupled with positions of previously confirmed companions in our sample 
(\citet{kirkpatrick01} and \citet{wilson01}) as a guide, and allowing for some scatter in colour. 
The 16 candidates companions in our selection box are shown as filled circles in this plot, and 
described in Table 1. The primary stars for each candidate are described in Table 2. Non-companions 
are found below our selection box and have J-K$\sim$1.1--1.3. The bottom left plot shows target distance 
against candidate companion separation, and it can be seen that all but one of the non companion candidates 
are at a large separation from a relatively nearby star. The characteristics of the non companions are thus 
entirely consistent with them being background late M and early L dwarfs in a relatively large space volume.

In order to estimate expected levels of contamination amongst our sample, we repeated our candidate 
identification procedures using the same target areas offset by 25000AU at the distance of each target, 
since we expect no companions at such large separations (see Figure 9 of \citet{burgasser03b}). The 
resulting candidates from these control fields are shown in the right two plots of Figure 3, and it 
can be seen that only 1 of these passes our photometric tests, demonstrating clearly that the vast 
majority of our candidate companions should be genuine.
\begin{figure*}
\includegraphics[width=120mm]{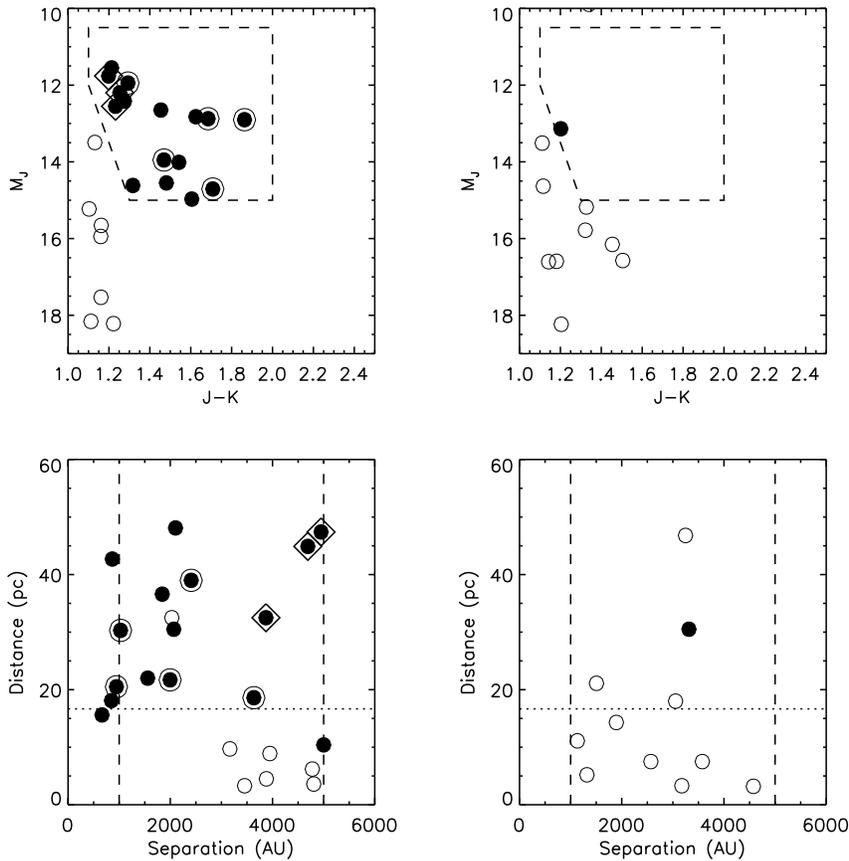}
\caption{M$_J$ verses J-K colour--magnitude diagrams (top) and distance--separation 
plots (bottom) for candidate wide L dwarf companions to Hipparcos stars (left) as well 
as those identified in a set of control fields, offset by 25000AU at the distance of 
each Hipparcos target. Filled symbols are candidates that pass our L dwarf photometric 
selection criteria (dashed box in top plots), assuming that the candidate is at 
the same distance as the primary star. Over-plotted circles and diamonds highlight 
companions confirmed through common proper motion from the literature and this work 
respectively (see text). Dotted lines in the bottom plots indicate the distance of 
16.67pc and the separation range 1000--5000AU.}
\end{figure*}

To further address the credence of our sample we have investigated available astrometry as well as 
additional photometry and spectroscopy. Candidate 8 has been spectroscopically measured as L3 (see 
DwarfArchives.org). Five of the candidates (3, 4, 5, 12 and 14) are already confirmed as common proper 
motion companions by \citet{kirkpatrick01} and \citet{wilson01}. Amongst the remaining objects five 
(candidates 2,6,8,9 and 10) are covered in the Sloan 4th data release (DR4) which provides optical 
measurements for three of these (candidates 2,8 and 10). All three have L dwarf colours compared 
to \citet{fan00} and \citet{hawley02}. Candidates 13 and 16 have optical counterparts in the 
SuperCOSMOS PossII and UKST I-band scans respectively, which also yield L dwarf like colours. 
The time base-line available from 2MASS to DR4 for candidate 2 is only 0.27 yr. However, longer 
baselines of 2.0yr, 6.1yr, 5.7yr and 14.9yr are available for candidates 8, 10, 13 and 16 respectively, 
from either 2MASS and DR4 or SuperCOSMOS and 2MASS. Taking into account the proper motions of the primary 
star candidates, we would expect motions of 1.1, 1.7, 2.0 and 2.0 arcseconds between epochs for these 
candidates respectively.

We used the geomap and geoxytran routines in Iraf (using 10--15 reference stars and an xy-shift/scale 
and rotation transformation) to transform either the Sloan pixel coordinates onto a 2MASS frame, or 
the 2MASS pixel coordinates onto a Schmidt frame. We found typical residuals of $\pm$0.2--0.3 arcseconds 
in our coordinate transforms, and estimate centroiding accuracies of $\pm$0.1--0.3 arcseconds depending 
on source brightness. As an additional test to estimate total astrometric uncertainty (including any 
chromatic effects in the astrometry of these very red sources), we derived the proper motion of 
candidates 8 and 10 using the full range of first epoch J, H and K images, and second epoch r', i' 
and z' images. This analysis suggests an over-all astrometric uncertainty of $\sim\pm$0.7 arcseconds, 
in reasonable agreement with our previous estimates.

The final proper motion determinations are shown in Figure 4. The candidate primary proper motions 
from Hipparcos are shown as circles, and the L dwarf candidate companions as squares. Each primary 
and candidate secondary are joined by a line for clarity. It can be seen that candidate 8 has large 
uncertainties, which is because the expected motion over the relatively short 2 yr baseline is 
comparable to the size of the astrometric uncertainties. Candidates 10, 13 and 16 show significant 
proper motion, consistent with that of the primary star in each case. These candidates are thus 
identified as common proper motion companions to HD 120005 (an FV spectroscopic binary), Gl 605 
(an M0 dwarf) and HD 216405 (a K1/K2 double star) respectively, and are listed in Table 1 as HD 
120005C, Gl 605B and HD 216405C. It is thus clear from the information presented in Figure 3 and 
Table 1 that our wide binary companion sample is robust.
\begin{figure}
\includegraphics[width=84mm]{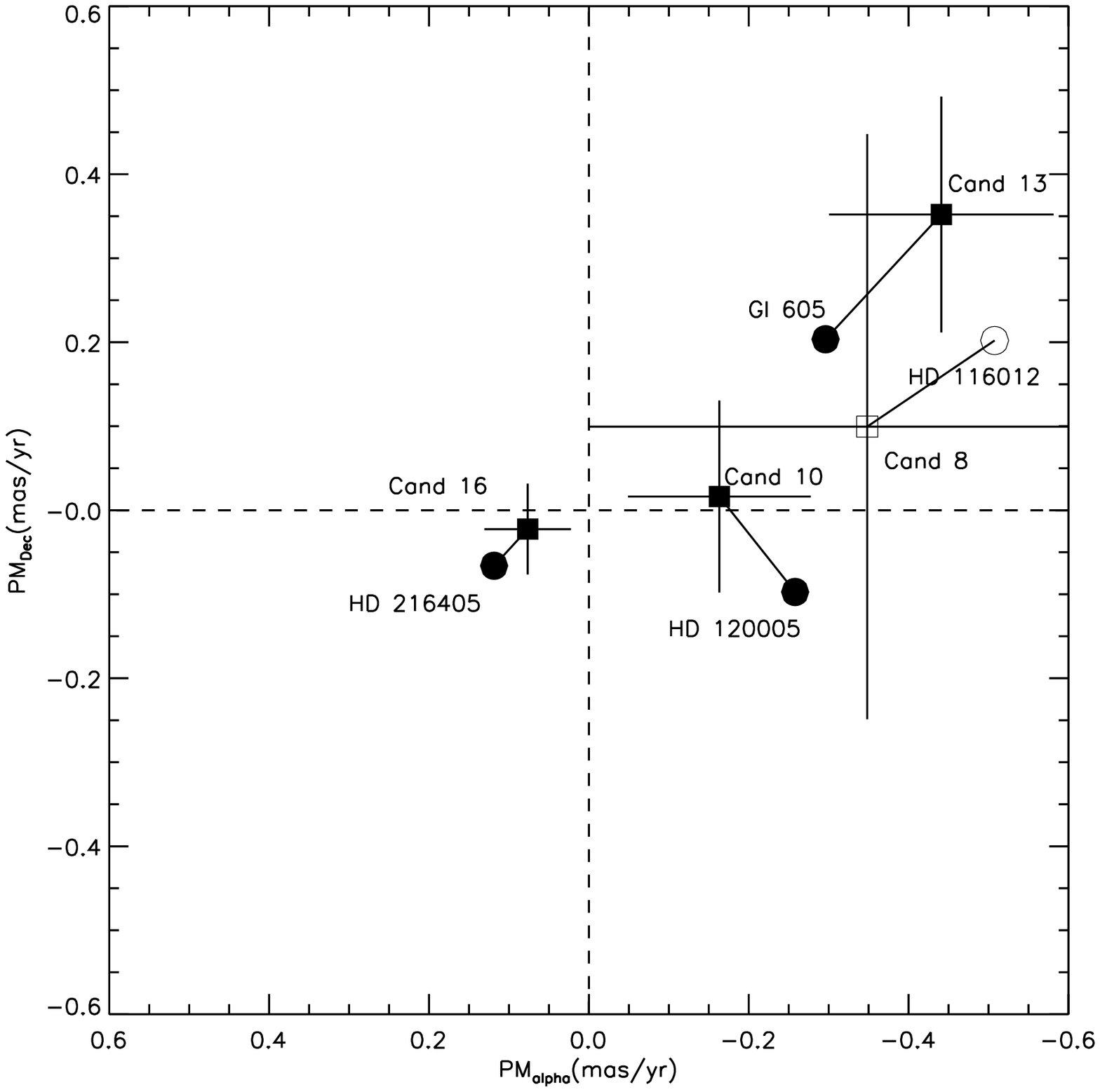}
\caption{Vector point diagram of the four candidate companions with useful epoch coverage (squares). 
The proper motions of the candidate primary stars (from Hipparcos) are shown as circles. The open symbols 
indicate a candidate L dwarf and primary where large measurement uncertainty results in an inconclusive 
result. The filled symbols indicate measured proper motions consistent with binary pairs. Hipparcos 
primaries and candidate companions have been joined by solid lines for clarity.}
\end{figure}

\begin{table*}
\caption{Wide companion L dwarf candidates.}
\centering
\begin{tabular}{|r|l|l|c|c|c|c|c|c|}
\hline
Cand & 2MASS source & Companion name & J & H & K & M$_J^d$ & J-K & a(AU) \\
     &              & (if confirmed) &&&&&&\\
     &              & and notes      &&&&&&\\
\hline
 1 & 2MASS J02124236+0341004 &                    & 14.70$\pm$0.03 & 13.91$\pm$0.04 & 13.38$\pm$0.04 & 14.62 & 1.32 & 5000 \\
 2 & 2MASS J08444996+5532121 & $^h$               & 14.70$\pm$0.04 & 13.95$\pm$0.04 & 13.49$\pm$0.02 & 11.55 & 1.21 &  865 \\
 3 & 2MASS J09121469+1459396 & Gl 337C$^b$        & 15.51$\pm$0.08 & 14.62$\pm$0.08 & 14.04$\pm$0.06 & 13.95 & 1.47 &  946 \\
 4 & 2MASS J10221489+4114266 & HD 89744B$^b$      & 14.90$\pm$0.04 & 14.02$\pm$0.03 & 13.61$\pm$0.04 & 11.95 & 1.29 & 2409 \\
 5 & 2MASS J11122567+3548131 & Gl 417B$^a$        & 14.58$\pm$0.03 & 13.50$\pm$0.03 & 12.72$\pm$0.03 & 12.90 & 1.86 & 1997 \\
 6 & 2MASS J11451802+0814414 &                    & 15.47$\star$   & 16.13$\star$   & 14.01$\pm$0.06 & 12.65 & 1.45 & 1842 \\
 7 & 2MASS J12201925+2636278 &                    & 15.83$\pm$0.16 & 13.97$\star$   & 14.55$\star$   & 12.42 & 1.28 & 2102 \\
 8 & 2MASS J13204427+0409045 & $^{e,i}$           & 15.25$\pm$0.05 & 14.30$\pm$0.03 & 13.62$\pm$0.05 & 12.82 & 1.62 & 2069 \\
 9 & 2MASS J13282546+1346023 &                    & 15.84$\pm$0.16 & 13.96$\star$   & 14.36$\star$   & 14.55 & 1.48 &  848 \\
10 & 2MASS J13460815+3055038 & HD 120005C$^{c,j}$ & 15.46$\pm$0.06 & 14.81$\pm$0.06 & 14.21$\pm$0.06 & 12.20 & 1.25 & 4691 \\
11 & 2MASS J13471545+1726426 &                    & 15.94$\pm$0.20 & 14.00$\star$   & 14.33$\star$   & 14.97 & 1.61 &  664 \\
12 & 2MASS J15232263+3014562 & Gl 584C$^a$        & 16.06$\pm$0.10 & 14.93$\pm$0.08 & 14.35$\pm$0.07 & 14.71 & 1.71 & 3635 \\
13 & 2MASS J15575569+5914232 & Gl 605B$^{c,f}$    & 14.32$\pm$0.03 & 13.61$\pm$0.04 & 13.12$\pm$0.03 & 11.76 & 1.20 & 3871 \\
14 & 2MASS J16202614-0416315 & Gl 618.1B$^b$      & 15.28$\pm$0.05 & 14.35$\pm$0.04 & 13.60$\pm$0.04 & 12.88 & 1.69 & 1027 \\
15 & 2MASS J17420515+7208002 &                    & 15.73$\star$   & 15.92$\star$   & 14.18$\pm$0.07 & 14.01 & 1.54 & 1558 \\
16 & 2MASS J22530539-3751335 & HD 216405C$^{c,g}$ & 15.93$\pm$0.08 & 15.35$\pm$0.09 & 14.70$\pm$0.09 & 12.55 & 1.23 & 4950 \\
\hline
\multicolumn{9}{|l|}{Notes:}\\
\multicolumn{9}{|l|}{$\star$: 95\% confidence upper limit.}\\
\multicolumn{9}{|l|}{$^a$Kirkpatrick et al. (2001). $^b$Wilson et al. (2001). $^c$This work. $^d$Assuming the same 
distance as the primary.}\\
\multicolumn{9}{|l|}{$^e$L3 (DwarfArchives.org). $^f$(R-K)$\sim$7. $^g$(I-K)$\sim$6. $^h$(i'-J)=4.3, (i'-z')=2.0.}\\
\multicolumn{9}{|l|}{$^i$(i'-J)=4.5, (r'-i')=2.4, (i'-z')=1.8. $^j$(i'-J)=4.2, (r'-i')=2.7, (i'-z')=1.9.}\\
\hline
\end{tabular}
\caption{Primary star properties for the candidate systems.}
\begin{tabular}{|r|l|l|c|c|c|c|c|}
\hline
Cand & Name & Hip & D(pc) & Spec Typ & V & M$_V$ & B-V \\
\hline
 1 & Gl 87         &  10279 & 10.4 & M1.5       & 10.04 & 9.95 & 1.43 \\
 2 & HD 74150      &  42919 & 42.7 & K0III-IV   &  8.91 & 5.76 & 0.80 \\
 3 & Gl 337AB      &  45170 & 20.5 & G9V/G9V    &  6.49 & 4.93 & 0.73 \\
 4 & HD 89744      &  50786 & 39.0 & F7IV-V     &  5.73 & 2.77 & 0.53 \\
 5 & Gl 417        &  54745 & 21.7 & G0V        &  6.41 & 4.73 & 0.60 \\
 6 & HD 102124     &  57328 & 36.6 & A4V        &  4.84 & 2.02 & 0.17 \\
 7 & HD 107325$^a$ &  60170 & 48.1 & K2III-IV   &  5.52 & 2.11 & 1.09 \\
 8 & HD 116012     &  65121 & 30.5 & K2V        &  8.58 & 6.16 & 0.94 \\
 9 & Gl 512.1      &  65721 & 18.1 & G5V        &  4.97 & 3.68 & 0.71 \\
10 & HD 120005$^b$ &  67195 & 44.9 & F5         &  6.51 & 3.25 & 0.49 \\
11 & Gl 527A$^a$   &  67275 & 15.6 & F6IV       &  4.50 & 3.53 & 0.51 \\
12 & Gl 584AB      &  75312 & 18.6 & G0V/G3V    &  4.99 & 3.64 & 0.58 \\
13 & Gl 605        &  78184 & 32.5 & M0         & 10.31 & 7.75 & 1.27 \\
14 & Gl 618.1$^a$  &  80053 & 30.3 & M0V        & 10.69 & 8.28 & 1.38 \\
15 & Gl 694.1A     &  86614 & 22.0 & F5IV-V     &  4.57 & 2.86 & 0.43 \\
16 & HD 216405$^c$ & 113010 & 47.4 & K1/K2V     &  9.36 & 5.98 & 0.88 \\
\hline
\multicolumn{7}{|l|}{Notes:}\\
\multicolumn{7}{|l|}{$^a$Variable star. $^b$Spectroscopic binary. $^c$Double/multiple star.}\\
\hline
\end{tabular}
\vfill
\end{table*}

With this greatly increased number of wide companions, we have been able to place significantly better 
constraints on the wide BD companion fraction than previously. Using the 14 candidates with distance 
$>$16.67pc for which our 2MASS search will be complete out to 5000AU, we have combined the value of M$_J$ 
for each companion with our J=16.1 search limit to determine the distance out to which we could have 
detected each companion. We then counted the number of Hipparcos targets out to this distance, establishing 
the number of stars (n$_{stars}$) around which a companion could have been detected. We then estimated the 
L dwarf companion fraction by determining the value of 1/$n_{stars}$ for each companion, and summing over all 
the companions. Finally, we made a small correction by subtracting off 1/$n_{stars}$ for the 1 object 
selected in our control fields. Our final L dwarf companion fraction is 2.7$^{+0.7}_{-0.5}$\%, where 
the uncertainties are $\pm$1-$\sigma$ assuming binomial statistics. This value is somewhat higher than 
the \citet{gizis01} estimate, although it is consistent to within their large uncertainties. If we assume 
that the fraction of BDs that can be detected as L dwarfs is 0.08 (for a companion MF with 
$\alpha\sim$1; see \citet{gizis01}), then we obtain a BD wide companion fraction of 34$^{+9}_{-6}$\%. 
This is somewhat higher than the wide stellar companion fraction over this separation range (10-15\%; 
\citet{duquennoy91}), suggesting that wide BD companions to main sequence stars could be quite common. 
However, the assumption that $\alpha\sim$1 represents an important source of uncertainty in this number. 
Nevertheless, in the remainder of this paper we will assume a BD wide companion fraction of 34$^{+9}_{-6}$\%.

Further constraints on the companion fraction and its mass distribution can be expected in the near 
future from the UKIDSS LAS. In way of demonstration we have simulated a population of BDs around 34\% 
of main sequence Hipparcos stars (estimating observable BD properties in the same manner as in Section 
2.1), and estimate that $\sim$50 L dwarf and 10--20 T dwarf wide companions may be identified. This 
population will provide improved statistics when estimating the companion fraction, and the spectral 
type and M$_J$ distributions of these companions will allow some constraints to be placed on the wide 
companion mass function. More detailed constraints would come from the population of $\sim$250 L dwarf 
and $\sim$150 T dwarf wide companions that could be available to similar photometric depth over the 
whole sky. The identification of these would require imaging around $\sim$10,000 Hipparcos stars, a 
task that would take over a hundred nights of 4m telescope time using a star-by-star approach. However, 
if facilities such as WFCAM, WirCam or Vista carry out all sky legacy surveys, this large population 
of wide L and T dwarf companions could be identified.

\section{Benchmark brown dwarfs as wide companions to subgiant stars}

Although the ages of main sequence stars are difficult to constrain with great accuracy, 
this is not the case for subgiants. Once a star has left the main sequence, and thick shell H burning 
occurs, it evolves almost horizontally across the HR diagram, before reaching the base of the giant 
branch. During this phase, its age can be accurately determined by comparison with evolutionary 
models. Theoretical predictions of subgiant evolution are sensitive to [M/H], but since subgiants 
are pre-dredge-up, [M/H] can be accurately measured, and ages inferred. The largest source of 
uncertainty in such evolutionary models is the extent of convective core over-shooting 
\citep{roxburgh89} that occurs for different masses, and this uncertainty is yielding to accurate 
observational constraints via the study of different aged open clusters (\citet{vandenberg04} and 
references therein). \citet{wilson01} has identified a wide L dwarf companion to an F7 IV-V primary 
using 2MASS, and it is clear from their figure 4 that this object gives a much better age calibration 
($\pm$30\% level) than the other 2 binaries reported. However, this star is only just leaving the 
main sequence, and we expect significantly better age calibration for fully fledged (class IV) 
subgiants. With [M/H] measured to 0.1 dex (eg. \citet{ibukiyama02}), and either distance known 
to 5\% or $\log{g}$ known to 0.1 dex, subgiant ages can be constrained to $\sim\pm$10\% accuracy 
\citep{thoren04}.

\subsection{Subgiant populations}

In order to identify the most comprehensive sample of currently available subgiants for our purposes, 
we have made use of the Hipparcos database. Figure 5a shows the M$_V$, B-V diagram of Hipparcos sources 
with V$<$13, $\pi$/$\sigma_{\pi}\ge$4 and B-V$>$0.6. Note that even if one has 25\% uncertainty on 
a Hipparcos subgiant parallax, one can subsequently derive an accurate spectroscopic distance, as 
demonstrated by the spectroscopic distance scale of \citet{fuhrmann98}, which agrees with Hipparcos 
parallaxes to an accuracy of 5\%. In order to select subgiants from this diagram, we have defined a 
colour-magnitude selection box using the solar metallicity isochrones of \citet{girardi00} as a guide 
(shown in Figure 5b). We chose the M$_V$ range 2.0--4.5 to include the majority of subgiants while 
avoiding the brightest ones (where glare would be more of a problem when attempting to image companions) 
and defined colour cuts to remove the majority of contamination from dwarfs and giant stars. The 
resulting subgiant selection box is indicated in both these Figures. For different [M/H] the isochrones 
change, becoming brighter and bluer for lower [M/H]. This means that we expect some contaminating 
sub-solar metallicity giants in the top right of the selection box. Giants will be undergoing dredge-up 
and [M/H] calibrations are thus not valid for these stars \citep{feltzing01}. We have therefore removed 
any objects that have been spectroscopically flagged as giant-like using data available in the Hipparcos 
database.
\begin{figure}
\includegraphics[width=105mm]{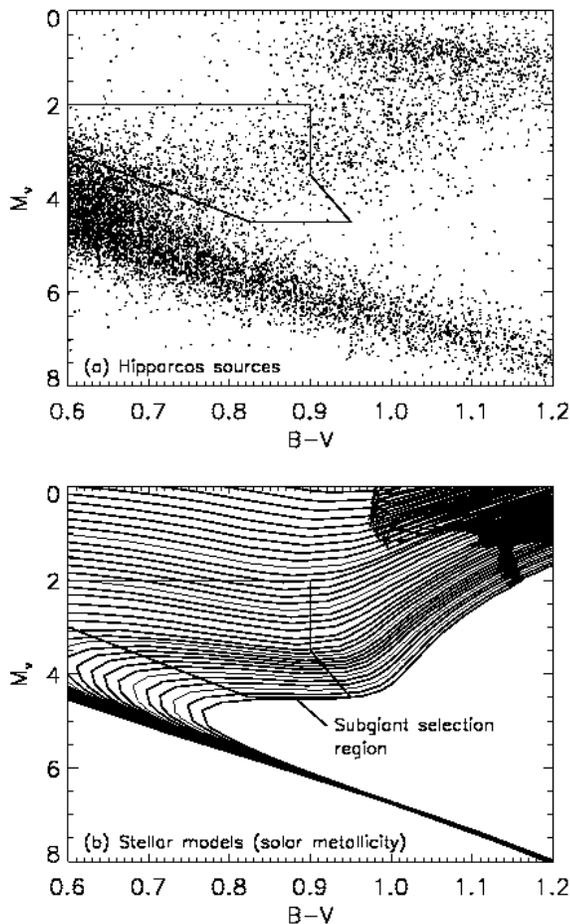}
\caption{(a) M$_V$, B-V diagram of Hipparcos stars with V$<$13 and 
$\pi$/$\sigma_{\pi}\ge$4. (b) Theoretical isochrones from \citet{girardi00} 
for solar [M/H]. Our subgiant colour--magnitude selection box is indicated 
in both plots.}
\end{figure}

Because we want to identify BDs around these subgiants, we would wish to avoid targets 
towards reddened regions, and we have thus used the reddening map of \citet{burnstein82} to identify 
and remove subgiants where galactic extinction is higher than A$_{\rm V}>$0.3 (E$_{\rm J-K}<$0.05). 
We would also wish to avoid over-crowded fields, where the extraction of accurate photometry will be 
problematic because of blended point-source-profiles. We therefore remove subgiants in directions where 
2MASS indicates there is $>$1.1 sources per square arcminute to J=15 (this translates into a typical 
nearest neighbour distance of $<$10 arcseconds to J=20; see Section 6.2). The over-crowded regions we 
avoid include a strip in the Galactic plane, as well as the LMC and SMC. Finally, we remove a small 
fraction of the subgiants with proper motions $<$40mas~yr$^{-1}$, since we would wish to follow-up 
and confirm candidate BD companions by measuring common proper motion. This could be done over a fairly 
short baseline of 1--2 yrs using adaptive optics imaging facilities such as NAOMI on the William Herschel 
Telescope, where a proper motion of 40mas/yr would produce a motion on the detector of 1 pixel per year. 
However, the majority of Hipparcos selected subgiants have proper motions of $\sim$50-200mas~yr$^{-1}$, 
and such proper motions could be measured using a more conventional approach. Finally, 
we impose a distance limit of 160pc on our subgiant sample, designed to facilitate the efficient discovery 
of approximately equal numbers of L and T dwarf benchmarks (see Section 6.2). Our selection criteria thus 
identify a target sample of 918 suitable subgiants.

\subsection{Simulating subgiant -- brown dwarf binary populations}

In order to simulate the properties of BD companions to subgiants, we must estimate an age 
distribution for the subgiant sample. Because location in the HR diagram depends on several 
factors ([M/H], mass and age), one cannot simply infer an age distribution from the M$_{\rm V}$, 
B-V diagram. Therefore, we estimated the age distribution using a simulated disk stellar population, 
for which we assumed a Salpeter MF, a birth rate history from \citet{rocha-pinto00}, a disk scale 
height-age relation from \citet{just03} and a metallicity distribution from \citet{edvardsson93}. 
Evolutionary tracks \citep{girardi00} were then used to derive the M$_{\rm V}$, B-V diagram for a volume 
limited sample (shown in Figure 6a), from which we selected a simulated subgiant sample using our 
colour-magnitude selection box to determine the expected age distribution (shown in Fig 6b).
\begin{figure}
\includegraphics[width=105mm]{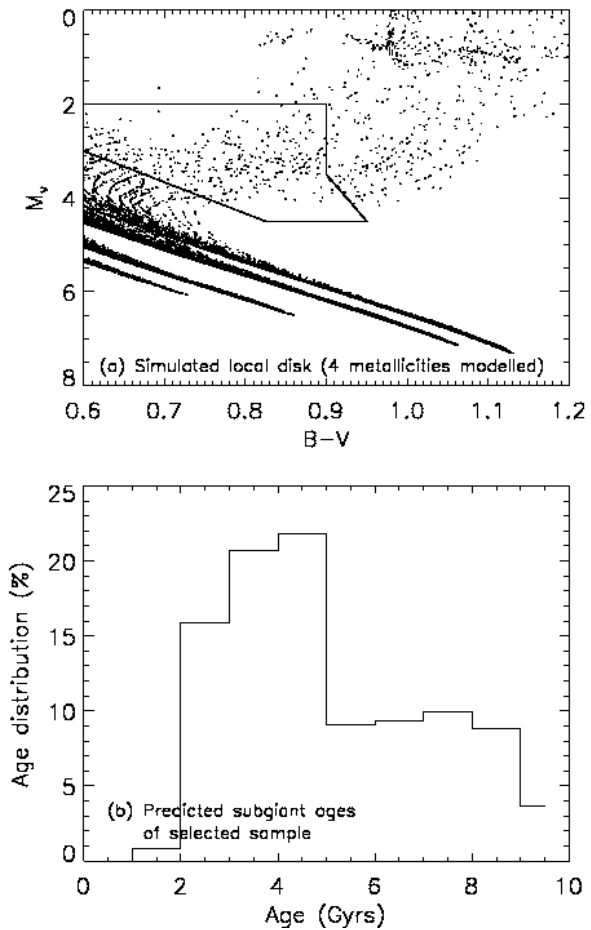}
\caption{(a) M$_V$, B-V diagram of our simulated local stellar population (see 
text). The main sequence appears as four distinct tracks because 4 distinct 
[M/H] values (0.2, 0.0, -0.35 and -0.65) were simulated (following the distribution 
of figure 14 from \citet{edvardsson93}. Our chosen colour magnitude selection box 
is also shown. (b) The age distribution for the simulated objects extracted from 
the subgiant selection box.}
\end{figure}

We randomly imposed our predicted age distribution on the subgiant sample, and added BD companions 
to 34\% of these, such that the BD masses follow an $\alpha$=1 mass function (see Section 5). Lyon 
group models (see Section 2.1) were then used to derive $T_{\rm eff}$, $g$ 
and M$_J$ from BD mass and age, and thus J magnitude at the distance of the subgiants.
\begin{figure}
\includegraphics[width=84mm]{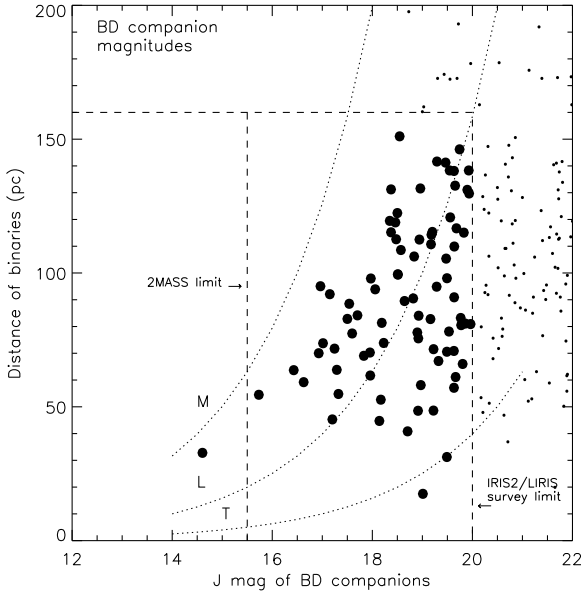}
\caption{The distance--magnitude distribution of the simulated 
BD companions to the Hipparcos subgiant population (see text). 
M, L and T dwarf divisions are shown with dotted lines. 
Distance--magnitude limits that should provide a good balance 
of L and T dwarf companions (see text) are indicated with dashed 
lines. The 2MASS limit is also indicated as a dashed line. The 
predicted $\sim$50 companions within these limits are highlighted 
by filled circles.}
\end{figure}

Figures 7 shows the resulting J magnitude--distance plot for the simulated companion BDs. M L and 
T dwarf divisions are shown with dotted lines. A photometric limit of J$\sim$20 will allow spectroscopic 
follow-up of benchmark BDs on 8-m telescopes (see Section 3). We chose a distance limit of 160pc 
to produce an evenly balanced number of L and T dwarf companions. This region of magnitude--distance 
space is shown in Figure 7, enclosed by dashed lines.

Our simulation predicts 80$^{+21}_{-14}$ wide companions in this region, approximately equally split 
between L and T dwarfs. The uncertainty associated with this number comes from the uncertainty of our 
estimated wide companion fraction (see Section 5).

\subsection{Finding subgiant brown dwarf binaries}

We have shown that $\sim$80$^{+21}_{-14}$ L and T dwarfs with J$<$20 could be expected around a sample of 918 
suitable subgiants within 160pc. 2MASS photometry is not deep enough to effectively probe this population 
(see Figure 5), and although the UKIDSS LAS survey reaches useful photometric depths, it covers only 10\% 
of the sky (and thus $\sim$10\% of the Hipparcos subgiants). Thus the discovery of this benchmark population 
requires an independent near infrared imaging campaign. Such imaging should be best done in the Y- J- and 
H-bands (where the Y filter covers the wavelength range $\sim$0.97--1.07 microns). With separations of 
$\sim$1000--5000AU and distances of $\sim$40-160 parsecs (see Fig 7), the angular separation of these L 
and T dwarfs from their subgiant primaries should vary from $\sim$6--125 arcseconds. Charge latency and 
cross-talk can be a problem when imaging on or near bright sources. But by locating the subgiants a few 
arcseconds off the edge of an infrared array, it should be possible (with two images -- either side of 
the subgiant) to image $\sim$95\% of the potential companion region, without imaging the subgiant directly.

Glare can also be an issue. The wings of the subgiant PSF will extend for several arcseconds. By assuming 
a Lorentzian PSF (with $\sim$1 arcsecond seeing) we have estimated the separation at which PSF brightness 
is at the same level as typical sky background (in the Y-band, where the sky brightness is lowest). By 
estimating these separations for our subgiant sample, where we average subgiant brightness as a function 
of distance, we have defined a minimum separation limit, above which the imaging of faint companions 
should not be affected by significantly enhanced sky background. This limit is shown in Figure 8 (dashed 
line), along with the distances and (randomly distributed) separations of the simulated companion population. 
It can be seen that very few of the companions should be significantly affected by an enhanced sky 
background. Figure 8 also shows the outer separation limits that would be imposed by NIR arrays the size 
of LIRIS on the WHT and IRIS2 on the AAT. We thus do not expect companions to be missed by such instruments.
\begin{figure}
\includegraphics[width=84mm]{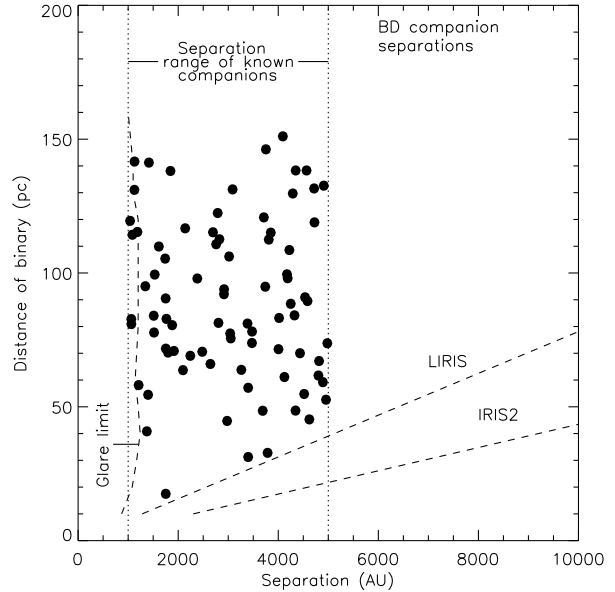}
\caption{The predicted separation--distance distribution for the 
subgiant BD companions. The dotted line shows the separation where 
we expect the PSF wings of typical subgiants to double the sky 
background in the Y-band. The dashed lines indicate the maximum 
separations covered by the large format NIR arrays IRIS2 on the 
AAT and LIRIS on the WHT.}
\end{figure}

Having imaged candidate wide companions, one needs to be confident that they are genuine, as opposed to 
random line of sight alignments. We have estimated the likely level of contamination from field objects 
by working out a contamination volume for each subgiant. We define this volume as that contained in a cone 
(apex at the observer) that points towards the target, has a cross sectional radius of 5000AU at the target 
distance, and covers a distance 63\%--158\% of the target distance. This distance range corresponds 
to a brightness range of $\pm$1 magnitude, and it should be possible to rule out field BDs in-front of, 
and more particularly beyond this distance range using colour magnitude information (as was done for the 
L dwarfs in Section 5). Within our sample contamination volume we have assumed 0.1 BDs per cubic parsec 
\citep{reid99} and thus expect 12 BDs to be contained in this volume. This number can be directly compared 
to the 312 BD companions (ie. a 34\% companion fraction) around the 918 targets, giving us a contamination 
fraction of $\sim$4\%. We therefore expect $\sim$3 field BDs to photometrically contaminate the $\sim$80 
genuine companions. This level of contamination is clearly low, and it is extremely unlikely that any of 
these contaminating objects would happen to share the proper motion of the primary subgiant. One should 
thus be able to confidently confirm such benchmark BDs from their photometry and proper motion.

\section{Benchmark brown dwarfs as wide companions to white dwarfs}

The basic physics entering white dwarf (WD) evolution has progressed significantly since the first detailed 
evolutionary calculations of \citet{lamb75}. Advances have been made on many fronts, including the 
convective and conductive opacities, the envelope equation of state, and the thermodynamics of the dense 
interior plasma including the effects of ion crystallization (see \citet{chabrier00a} and references 
therein). Furthermore, hot WD atmospheres of pure hydrogen can be well modelled \citep{hubeny95}, and 
the $T_{\rm eff}$ and $g$ measured accurately by fitting synthetic spectra to Balmer lines in the optical 
(e.g. \citet{dobbie05}; \citet{claver01}). WD cooling ages can thus be well determined using the measured 
$T_{\rm eff}$ and $g$ (and an assumed mass--radius relation) and an evolutionary model. It is not, however, 
possible to establish the [M/H] of a WD progenitor from observations of the WD, since the WD surface 
composition has little bearing on the progenitor composition.

WD--BD binaries have been the subject of numerous searches in the past. These searches have generally 
focused on finding resolvable, but relatively close companions. For example, GD~165B was the first of 
its type discovered, consisting of a WD-L dwarf (probably not quite substellar) binary with a separation 
of $\sim$150AU \citep{zuckerman92}. We do not expect many BD companions to solar type stars at such 
separations (see Section 5), although less is known about BD companions to higher mass stars (WD 
progenitors). Note also that when a star becomes a WD, its undergoes mass loss, and during this process 
any wide companions will spiral out to greater separation \citep{burleigh02}. When, for example, a 
3M$_{\odot}$ star becomes a WD, we expect its mass to decrease by a factor of $\sim$4 \citep{williams04}. 
Companions separated by 1000--5000AU would spiral out to 4000--20000AU when the star becomes a WD. 
While some of these binaries may be disrupted due to stellar encounters, we would expect many containing 
higher mass WDs to remain intact (see figure 9 of \citet{burgasser03b}). Previous NIR imaging searches 
have not effectively probed this very wide separation range -- eg. \citet{farihi03} searched for faint 
companions to WDs within 90 arcsecond separations ($<$4500AU at the typical 50pc distances of their 
sample). Wider separation ranges must be searched to identify the very wide binaries in question.

The most important source of uncertainty in the ages of any companions to WDs will generally 
be the unknown lifetime of the main sequence progenitor star. However, this uncertainty can be minimised 
by selecting WDs with small progenitor lifetimes. It is known, for example, from the study of open cluster 
WDs that there is a relationship between the progenitor mass and the WD mass (the so 
called 'initial-mass-final-mass-relation' or IMFMR), except for a small number of cases where one most 
likely has a magnetic WD (mass loss may be inhibited by a strong magnetic field). The IMFMR is 
shown in figure 6 of \citet{williams04}, and demonstrates that if a WD mass is $>$0.7M$_{\odot}$, one can 
place a lower limit on the progenitor mass, and hence an upper limit on the progenitor lifetime, using 
stellar models.

In order to define good age calibrating WDs, we require the progenitor life-time to be no more than 
10\% of the WD cooling age. This way, the WD cooling age will be an accurate measure of the total binary 
age, and hence the age of any BD companion. We show this age calibration criterion on a WD mass--age 
diagram in Figure 9a. It can be seen that all WD age calibrators have mass$>$0.7M$_{\odot}$, and that 
the minimum mass increases for ages $<$2 Gyr. Figure 9b shows the age calibrating criterion transformed 
onto a $T_{\rm eff}$--$\log{g}$ diagram, where we overplot WDs from the Sloan first data release 
\citep{kleinman04}. Only $\sim$15\% of these Sloan WDs are good age calibrators, and many of these 
will have $T_{\rm eff}$ from 7500--10000K (u'-g'$\sim$0.3--0.6, g'-r'$\sim$-0.2--0.3). We thus expect 
some overlap with the colours of stars (see figure 1 from \citet{smolcic04}). Proper motion analysis 
will thus be an important tool in the identification of age calibrating WDs, allowing selection based 
on reduced proper motion diagrams (e.g. \citet{knox99}, \citet{munn04}).
\begin{figure}
\includegraphics[width=84mm]{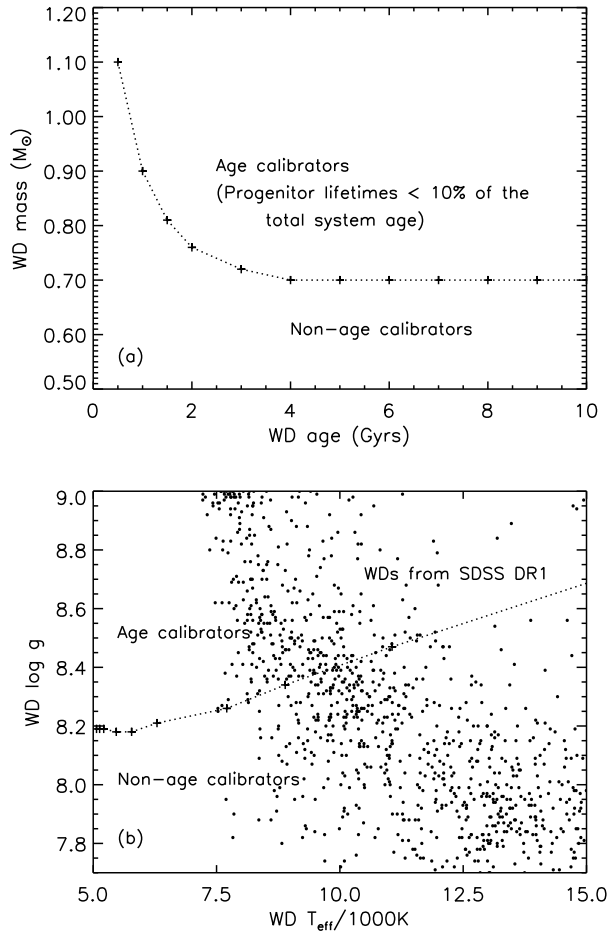}
\caption{(a) The mass--age plane for WDs. The dotted line indicates 
where the good age calibrating WDs lie. Above this dotted line, 
one can place mass constraints on the WD progenitor star that in 
turn limit its main sequence lifetime to be less than 10\% of the 
WD cooling age. (b) The good age calibration separator (dotted line 
in (a)) transformed into $\log{g}$--$T_{\rm eff}$ space. Sloan WDs 
from \citet{kleinman04} are over-plotted.}
\end{figure}

\subsection{Simulating white dwarf -- brown dwarf binary populations}

In order to create a synthetic WD disk population, we initially follow the approach of 
\citet{schroeder04}. We define a number--distance relation such that n$\propto$d$^3$ (normalised 
to 37 WDs within 13pc) out to 50pc. Beyond 50pc we assume that n$\propto$d$^{2.7}$ to account for 
reduced number densities as one approaches the average disk scale height of $\sim$250pc. We define 
our WD mass distribution using Figure 2 from S04. A complete WD age distribution will be complicated 
by the time-scales for stellar evolution. However, since we will preselect only WDs that are good 
age calibrators (ie. with relatively short progenitor life-times), we can make the simplifying 
assumption that our WD age distribution is the same as the stellar age distribution. We thus assume 
a WD birth rate identical to the stellar birth rate of \citet{rocha-pinto00}, and derive an age 
distribution by correcting for an age dependent disk scale height, as previously. Synthetic WD 
properties (luminosity, $T_{\rm eff}$ and $g$) were derived from mass and age, using equation (1) 
of S04, and the mass-radius relation of \citet{panei00}.

Simulated WD photometric properties were then determined using a combination of colour--$T_{\rm eff}$ 
and BC--$T_{\rm eff}$ information from models \citep{chabrier00b} and observation \citep{kleinman04}. 
Photometry was transformed onto the Sloan and photographic systems as required using \citet{bessell86} 
and \citet{smith02}. Simulated proper motions were also derived by imposing a tangential velocity 
(V$_{tan}$) distribution on our population. We assumed an old disk velocity ellipsoid in the UV plane 
(centred at V,U=-35,0kms$^{-1}$ and with a velocity dispersion of 45kms$^{-1}$) for ages $>$1Gyr, and 
a young disk velocity ellipsoid (U=-20--50kms$^{-1}$, V=-30--0kms$^{-1}$) for younger ages. Total proper 
motions were then determined from V$_{tan}$ and distance. Motion in the UV plane alone will be entirely 
appropriate when looking in the Galactic cap (i.e. Sloan and UKIDSS LAS), but at lower galactic latitudes 
we would expect a W component in V$_{tan}$. However, since it is the size of the proper motions that 
concerns us (their detectability), estimating the V$_{tan}$ distribution across the whole 
sky from the UV plane alone should be a good approximation.

We then selected only the good WD age calibrators, using the $T_{\rm eff}$--$g$ criteria from Figure 9, 
and randomly added wide BD companions to 34\% of these, in line with our derived wide companion fraction 
(see Section 5). An $\alpha$=1 mass function was imposed on the companion population, and the ages of 
the BD companions were set to be the same as their WD primaries. BD brightness was then determined from 
mass and age using the Lyon Group models, and converted to apparent magnitudes using the distance of 
each binary system.

\subsection{Finding white dwarf -- brown dwarf binaries in large scale surveys}

To determine the number of wide benchmark WD-BD binaries that could be found in large scale surveys, 
we have extracted photometrically and astrometrically limited samples from our simulated WD-BD population. 
In the optical, where WDs can be best detected, we consider Schmidt plate surveys covering the 
whole sky in the B- and R-bands, as well as Sloan which will cover $\sim$25\% of the sky in five 
bands. SuperCOSMOS for example produces accurate photometry with proper motions uncertainties of 
$\pm$10mas/yr down to B=19, R=18 \citep{hambly01}. With Sloan, one can probe more deeply with accurate 
proper motions. \citet{munn04} combines USNO-B with Sloan to produce a catalogue of absolute proper 
motions (re-calibrated using Sloan astrometry of galaxies) with improved statistical errors of $\sim$4mas/yr, 
due in part to the additional Sloan epoch. This catalogue is 90\% complete to g'=19.7. BDs may be 
selected from Sloan by their large i'-z' colour, where one is limited by i'-band brightness. In the NIR, 
we consider BDs detected in the 2MASS All Sky Survey, and in the UKIDSS LAS (LAS photometric limits 
were described in Section 2).

We consider three main survey combinations in which to find benchmark WD-BD binaries. Firstly, one could 
use 2MASS to find BDs and Schmidt plate photometry and proper motions to identify WDs across the majority 
of the sky (ignoring regions in the plane and the Magellanic clouds; see Section 6.1). Secondly, one 
could probe slightly deeper using Sloan+USNO-B to find WDs, and Sloan i'-z' colours to identify BDs 
(Sloan is sensitive to early L dwarfs out to greater distances than 2MASS). Finally, one could combine 
the UKIDSS LAS with Sloan+USNO-B, to reach the faintest NIR and optical limits for 10\% of the sky. Using 
the appropriate magnitude limits, requiring that simulated proper motions are $>5\times\sigma_{pm}$ (where 
$\sigma_{pm}$=10mas/yr and 4mas/yr for SuperCOSMOS and USNOB+Sloan respectively), and accounting for the 
fractions of sky covered, we extracted benchmark WD--BD binaries from our simulated population. Our results 
suggest that one could find $\sim$9 benchmark systems using 2MASS/Schmidt plate data, $\sim$6 using Sloan 
by itself, and $\sim$50$^{+13}_{-10}$ benchmark systems from UKIDSS LAS combined with Sloan+USNO-B, where 
the uncertainties are associated with those of the wide companion fraction (Section 5).

Note that in practise one would also expect to find $\sim$5 times as many non benchmark WD-BD binaries. 
For example, there could be $\sim$300 WD-BD binaries in total amongst the $\sim$4000 WDs expected in the 
UKIDSS LAS and USNOB+Sloan combination, with only $\sim$50 of these systems containing age calibrating 
WDs and benchmark BDs.

\begin{figure}
\includegraphics[width=84mm]{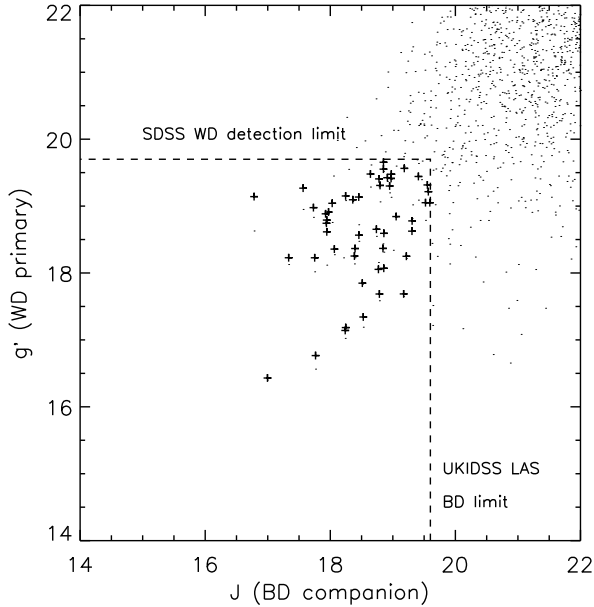}
\caption{The simulated WD--BD wide binary population from 10\% of the sky. Sloan g'-magnitudes for 
WDs are plotted against BD J-magnitudes. Sloan and UKIDSS LAS photometric limits are indicated with 
dashed lines. Identifiable benchmark binaries in our simulated population are indicated with crosses.}
\end{figure}
Figure 10 shows BD magnitude against WD magnitude for our simulated wide benchmark binaries in the LAS/Sloan 
selection. Note that this is for 10\% of the sky, and although there are no systems where the BD companion 
has J$<$16, the other survey combinations cover a larger area of sky, and are able to identify some brighter 
BD companions. The main limiting factor for these other combinations is the detection of BDs using 2MASS J 
or Sloan i'. The significantly greater photometric depth of the LAS provides sensitivity to a much larger 
number of benchmark systems, and it can be seen that the optical and NIR depths of Sloan and the LAS are 
quite well matched for this purpose.

We have estimated likely levels of contamination in the simulated LAS/Sloan+USNO-B population as we did for 
the subgiant companions in Section 6.3. However, note that WD companions are expected to be in wider orbits, 
so we assume a cone with a 20000AU radius at the distance of the target. Using the same approach as in 
Section 6.3 we estimate that $\sim$400 contaminating field objects will have photometry consistent with 
companionship. This number is comparable with the expected number of WD-BD binaries (cf. $\sim$300). 
Clearly such contamination is more of an issue for these systems than it was for subgiant companions. 
However, common proper motion should still be a very effective way of confirming genuine binary companions.

\subsection{Confirming white dwarf -- brown dwarf benchmark systems}

Having established companionship through common proper motion, optical spectroscopy of the WDs on 4--8m 
class telescopes, will allow their $T_{\rm eff}$ and $\log{g}$ to be measured. This will establish if 
they are good age calibrators (according to Figure 9b), and allow their cooling ages to be determined. 
Our simulation suggests that the WD--BD systems will have distances ranging from $\sim$25--300pc, so 
although parallax distances could be derived for the closer binaries, alternative WD distance constraints 
will be required for many, if the BD is to be a benchmark object. Such distance constraints can also come 
from the $T_{\rm eff}$ and $\log{g}$ measurements when combined with WD evolutionary models, since together 
these allow the WD mass and radius (and thus luminosity) to be measured. Such spectroscopically determine 
properties should be quite robust, as demonstrated by \citet{claver01}, who studied populations of open 
cluster WDs and showed that the spectroscopic masses agree very closely with those derived via an assumed 
cluster distance, and with those derived via gravitational redshifts. With a WD cooling age and 
spectroscopic or parallax distance for a binary, the BD $T_{\rm eff}$ and $\log{g}$ can be derived 
(see Section 3), creating a benchmark BD.

\subsection{Spectral types of wide brown dwarf companions to white dwarfs}

\begin{figure}
\includegraphics[width=84mm]{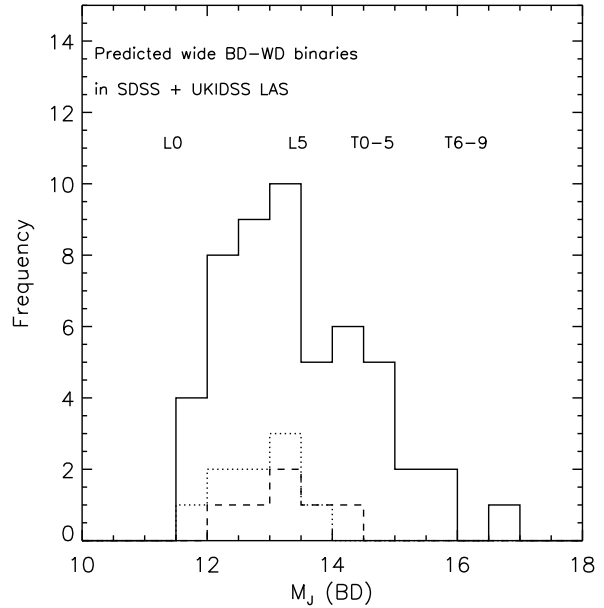}
\caption{M$_J$ distribution of the simulated wide BD companions to WDs. 
Those detectable in Sloan combined with the UKIDSS LAS are shown by the 
solid histogram. Those detectable in 2MASS combined with Schmidt plate 
coverage, and by Sloan alone, are shown by the dotted line and dashed 
line histograms respectively. Approximate spectral class locations are 
indicated at the top of the plot.}
\end{figure}
Figure 11 shows the M$_J$ frequency distribution of the simulated BD companions to WDs from the LAS 
combined with Sloan+USNO-B simulation (solid histogram). Approximate spectral class locations are 
indicated near the top of the plot. Also plotted are the frequency distributions for the simulated 
systems from 2MASS combined with SuperCOSMOS (dotted histogram) and Sloan by itself (dashed histogram). 
The low numbers of BD companions detectable in the absence of the UKIDSS LAS are mostly L dwarfs. 
2MASS/SuperCOSMOS should identify slight more systems, but Sloan may probe to slightly later BDs. 
However, the simulated population from LAS combined with Sloan covers the M$_J$=11.5--16 range, 
corresponding to L--late T. The LAS is clearly vital to both increase the number of WD-BD benchmark 
systems as well as extending the spectral type range to the later T dwarfs.

\section{The benchmark brown dwarfs in mass-age space}

\begin{figure}
\includegraphics[width=84mm]{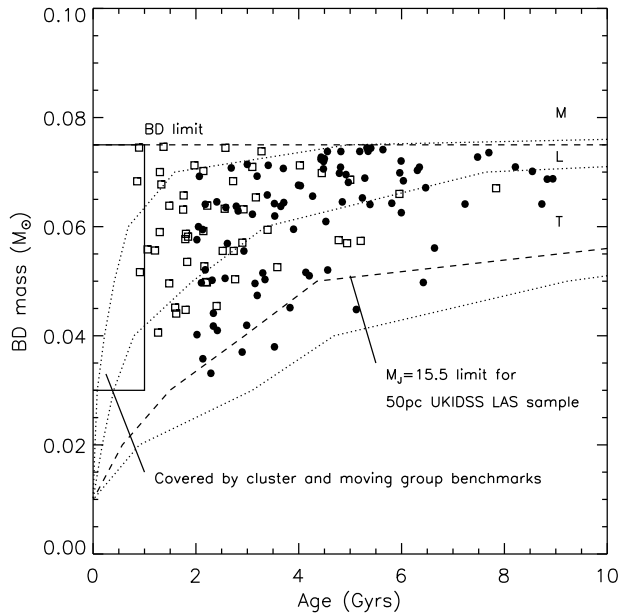}
\caption{The mass--age distribution of the benchmark BDs discussed in this 
work. Estimated spectral class divisions are shown with dotted lines. The 50pc 
limit for $\sim$T7 dwarfs in the UKIDSS LAS is shown as a dashed line. Together 
with the BD limit (also a dashed line) these enclose the region in which we 
expect the UKIDSS LAS to measure the BD PDMF and formation history. The mass--age 
region covered by young clusters and moving groups is enclosed by the solid line 
rectangle. Predicted benchmark BDs as wide companions to subgiant stars, and as 
wide companions to high mass WDs are shown as filled circles and open squares 
respectively.}
\end{figure}
Figure 12 shows the predicted mass-age distribution of the different benchmark populations discussed 
in this work (annotation is the same as in Figure 1a). The mass-age region in which UKIDSS LAS 
could accurately measure the BD IMF and formation history is enclosed by the dashed lines (see 
Section 2). The mass-age region that we expect to be covered by cluster and moving group benchmarks 
is indicated by the solid line rectangular box, representing 0.03--0.075M$_{\odot}$ and $<$1 Gyr.

The simulated wide binary BD companions to subgiants and high mass WDs are shown as filled circles and open 
squares respectively. The two types of binary BD have significant overlap in mass-age space. However, note 
that there are particular mass-age regions that are primarily covered only by one type of binary companion. 
Due to the deeper coverage of our simulated subgiant survey, the subgiant companions populate the late T 
dwarf region more effectively than the WD companions found in the LAS (see also Figure 7 and Figure 11). 
This means that this benchmark population is particularly important in the context of deriving the BD 
formation history, as well as the lower mass end of the intermediate age disk IMF (eg. 0.04--0.06M$_{\odot}$, 
2--4 Gyr). However, the subgiant benchmarks do not cover the age range $<$2Gyr (see also Figure 6b). 
The benchmark WD companions are not very common for ages $>$5Gyrs. However, they do cover the 1--2 Gyr 
age range.

It is also important to consider the benchmark BD number density within $T_{\rm eff}$--$g$--[M/H] space. 
For the older systems (4--10 Gyr), the relatively small mass range covered ($\sim$0.055--0.075
M$_{\odot}$) results in a small $g$ range ($\log{g}\simeq$5.3--5.4), but a large $T_{\rm eff}$ range 
($\sim$2000--1000K). In this part of the mass--age diagram, it is important to thoroughly populate 
the $T_{\rm eff}$--[M/H] plane as much as possible, using the subgiant companion benchmarks. We expect 
the [M/H] distribution of these benchmarks to follow the [M/H]--age distribution from \citet{edvardsson93} 
(i.e. the [M/H] distribution that we assumed for the subgiant primaries in our simulation). As shown in 
Figure 14 of \citet{edvardsson93}, on average there is a slight preference for lower [M/H] systems at 
older ages. However the spread in [M/H] at all ages is significantly larger than this trend, and for 
4--10 Gyr ages [M/H] is quite uniformally spread from $\sim$-0.7--+0.2 (i.e. a spread of $\sim$1 dex). 
Our simulations predict $\sim$40 subgiant companion benchmarks in this age range, and with the expected 
$T_{\rm eff}$ and [M/H] distributions we thus expect $\sim$1 benchmark BD per $\Delta$$T_{\rm eff}$=200K 
$\Delta$[M/H]=0.1 dex.

For the younger BDs (1--4 Gyr), the larger mass range covered will result in a larger $g$ range 
($\log{g}\simeq$5--5.3) as well as a large $T_{\rm eff}$ range ($\sim$2000--1000K). It is thus 
important to more fully populate this part of the mass-age plane, and the $\sim$80 simulated 
subgiant and WD companion benchmarks should give $\sim$1 benchmark BD per $\Delta$$T_{\rm eff}$=100K, 
$\Delta g$=0.04 dex.

\section{Conclusions}

In this paper we have discussed the current prospects for identifying field BD populations, 
and examined their likely mass, age and [M/H] distribution. We also consider the idea that 
it may be possible to constrain their properties using a combination of spectroscopically 
determined $T_{\rm eff}$, $g$ and [M/H] combined with distance constrains and an evolutionary 
model. We suggest that the spectroscopic calibration of these properties might be realised 
via the study of populations of BDs whose properties are well constrained by independent 
means -- so called benchmark BDs. We consider the different types of benchmark BDs that 
could be discovered, the number of benchmarks that we may expect to find in the near future, 
and the range of properties that we can expect them to have. Our conclusions may be summarised as 
follows:

\begin{itemize}

\item{The UKIDSS LAS should discover large numbers of field BDs covering a substellar mass range 
down to 0.03M$_{\odot}$ (for ages $<$1.5Gyr), and an age range out to 10 Gyr (for masses from 0.055
--0.075M$_{\odot}$).}

\item{The best sources of young ($<$1 Gyr) benchmark BDs should be in open clusters and moving 
groups. Some benchmarks of this type are known already, but many more are expected in the next 
few years from the UKIDSS GCS survey. These clusters all have [M/H]=-0.1--0.2 however, and the 
[M/H] of additional young clusters must be measured in order to provide hunting grounds for the 
most metal rich and metal poor young benchmark BDs.}

\item{The best sources of older ($>$1 Gyr) benchmark BDs should be as wide companions to subgiant 
stars and high-mass ($>$0.7M$_{\odot}$) WDs. BD $T_{\rm eff}$, $g$ and [M/H] can be accurately 
constrained by association with subgiant companions, and high mass WD companions may be used to 
constrain BD $T_{\rm eff}$ and $g$.}

\item{A NIR survey around $\sim$900 available Hipparcos subgiants could find $\sim$80$^{+21}_{-14}$ 
benchmark BDs. Such benchmark objects will be particularly useful for revealing spectral sensitivities 
to [M/H], and to the $T_{\rm eff}$ and $g$ of older BDs.}

\item{The combination of the UKIDSS LAS and Sloan surveys could find $\sim$50$^{+13}_{-10}$ BD 
companions to high-mass WDs. These benchmark BDs should be particularly useful for studying the 
spectral sensitivities to $T_{\rm eff}$ and $g$ of BDs in the 1--2 Gyr age range.}

\item{Together, the available benchmark populations cover the mass--age range in which the PDMF and 
formation history can be measured. Also, the [M/H] distribution of the benchmark BDs seems likely 
to encompass the [M/H] range expected amongst field BDs, accepting the caveat about the young cluster 
benchmarks. BD $T_{\rm eff}$--$g$--[M/H] space should be well populated by these benchmark objects, 
which could thus provide a grid of fiducial benchmark BDs for spectroscopic study.}

\end{itemize}

The identification of these benchmark populations could thus provide a foundation to allow the 
UKIDSS LAS to accurately probe the BD PDMF down to 0.030M$_{\odot}$, and the substellar 
(0.055--0.075M$_{\odot}$) formation history from 0--10 Gyr.

\section*{Acknowledgements}

We acknowledge support from PPARC for this work, in particular for their support of DP, PL, TK \& SF. 
This research has made use of data obtained from the Leicester Database and Archive Service at the 
Department of Physics and Astronomy, Leicester University, UK.

\clearpage

\bsp

\label{lastpage}

\end{document}